\documentclass{ar-1col}

\usepackage{url,natbib,amsmath,amssymb}

\jname{Xxxx. Xxx. Xxx. Xxx.}
\jvol{AA}
\jyear{YYYY}
\doi{10.1146/((please add article doi))}
\begin{document}

\markboth{Owen, James E.}{Atmospheric escape from exoplanets}

\title{Atmospheric Escape and the Evolution of Close-in Exoplanets}

\author{James E. Owen
\affil{Astrophysics Group, Imperial College London, Blackett Laboratory, Prince Consort Road, London SW7 2AZ, UK; email: james.owen@imperial.ac.uk}}

\begin{abstract}
Exoplanets with substantial Hydrogen/Helium atmospheres have been discovered in abundance, many residing extremely close to their parent stars. The extreme irradiation levels these atmospheres experience causes them to undergo hydrodynamic atmospheric escape. Ongoing atmospheric escape has been observed to be occurring in a few nearby exoplanet systems through transit spectroscopy both for hot Jupiters and lower-mass super-Earths/mini-Neptunes. Detailed hydrodynamic calculations that incorporate radiative transfer and ionization chemistry are now common in one-dimensional models, and multi-dimensional calculations that incorporate magnetic-fields and interactions with the interstellar environment are cutting edge. However, there remains very limited comparison between simulations and observations. While hot Jupiters experience atmospheric escape, the mass-loss rates are not high enough to affect their evolution. However, for lower mass planets atmospheric escape drives and controls their evolution, sculpting the exoplanet population we observe today. 
\end{abstract}

\maketitle

\tableofcontents


\section{INTRODUCTION}

The last 25 years have seen remarkable growth in the number of detected exoplanets. The very nature of our detection methods means we are often biased towards finding planets close to their host stars. The early discoveries were dominated by planets with approximately the same mass and radius as Jupiter, but with orbital periods in the range 1-10 days. These ``hot-Jupiters'' were our first glimpse of highly irradiated exoplanets and the importance of atmospheric escape\footnote{A note on terminology: the literature is littered with different terms for the escape of gas from a planet's atmosphere due to heating. Here we use the umbrella term of atmospheric escape, but photoevaporation, evaporation and blow-off are commonly used. } processes were discussed in the original discovery papers for the first hot Jupiter  \citep[51-Peg.\,b,][]{Mayor1995,Burrows1995}. While hot-Jupiters are rare, with an occurrence rate around 1$\%$ \citep[e.g.][]{Cumming2008,Howard2010,Howard2012}, their discovery early in the history of exoplanet science and their relative ease of follow-up observations meant that a large fraction of the early theory and observations of atmospheric escape from exoplanets was focused on hot Jupiters.

At the beginning of this decade NASA's {\it Kepler} mission began to reveal that even smaller and lower mass planets were present at short periods \citep[e.g.][]{Borucki2011,Thompson2018}. Calculations of the occurrence rates of these ``super-Earths'' and ``mini-Neptunes'' indicated that these small (1-4~R$_\oplus$), low-mass ($\lesssim 20$~M$_\oplus$)  planets are incredibly common, with many FGK stars hosting at least one of these planets with an orbital period shorter than 100~days \citep[e.g.][]{Fressin2013} and they appear to be even more common around M stars \citep[e.g][]{Dressing2013}. These close-in, small planets are the most common type of known exoplanet today. 

While we know hot-Jupiters' atmospheres are dominated by a mixture of predominantly Hydrogen and Helium, it is still unclear what constitutes the atmospheric composition of super-Earths and mini-Neptunes. When we can measure both a planet's radius and mass, the planet's density can be compared to theoretical structure models. In some cases, this comparison reveals that the planet is consistent with being entirely solid \citep[e.g.][]{Dressing2015}; in other cases the planet has such a low density that it needs to have a significant (Earth radii or larger) atmosphere composed of Hydrogen/Helium gas \citep[e.g.][]{Wu2013,Hadden2014,Weiss2014,Jontof-Hutter2016}. However, at intermediate densities, a broad range of atmospheric compositions are possible including water-rich and Hydrogen-Helium rich \citep[e.g.][]{Rogers2010a,Rogers2010b}.

The presence of exoplanets with volatile atmospheres at such short orbital periods raises the question as to whether such atmospheres are stable \citep[e.g.][]{Koskinen2007}. For Hydrogen dominated atmospheres, the high UV fluxes close to the star dissociates the molecular Hydrogen, resulting in the upper regions being dominated by atomic Hydrogen \citep[e.g.][]{Yelle2004,Munoz2007}. In these atomic regions heating typically results in gas temperatures of order 5,000-10,000~K. At such high temperatures, the upper atmospheres of close-in exoplanets are weakly bound, and the gas can escape the planet's gravity. In many cases, and especially for small, low-mass planets the total time-integrated high-energy flux (high-energy exposure) a planet receives over its billion year lifetime is a significant fraction of its gravitational binding energy \citep[e.g.][]{Lecavelier2007}. This available energy means that unlike Solar-System planets, where atmospheric escape is vital for shaping the chemical evolution of a planet's atmosphere \citep[e.g.][]{Lammer2008}, in the context of close-in exoplanets, atmospheric escape can affect the evolution of a planet's bulk composition. As we shall see, it has become clear that atmospheric escape is a key evolutionary driver in the evolution of many close-in planets. 

This review is focused on escape from exoplanets whose atmospheres are dominated by Hydrogen and Helium building on previous reviews by \citet{Yelle2008} and \citet{Tian2015}; however, in Section~\ref{sec:future} we shall discuss how this work can and must be connected to atmospheric escape calculations of atmospheres that are dominated by heavy elements. We also note that atmospheric escape from ultra-short-period planets that are close enough to have their rock surfaces vaporise and escape has been observed and studied theoretically. This topic has been reviewed recently in \citet{vanLieshout2017}.

This review is organised as follows: in Section~\ref{sec:obs} we discuss the observational probes of ongoing atmospheric escape from exoplanets; in Section~\ref{sec:theory} we discuss the status of our theoretical understanding and modelling efforts; in Section~\ref{sec:evolve} we discuss the evolutionary impact of atmospheric escape and it's imprints within the exoplanet population and finally in Section~\ref{sec:future} we discuss some of the open questions and future problems that need to be tackled.

\section{OBSERVATIONS OF ATMOSPHERIC ESCAPE}\label{sec:obs}
Remarkably, observations of atmospheric escape emerged relatively shortly after the detection of the first transiting planets. The observational technique is essentially the same as the transit method: by looking for the absorption of stellar light as the planet transits in front of the star, one can measure the ratio of the area of any obscuring, escaping atmosphere to the area of the stellar disc. To confirm the presence of atmospheric escape one needs to show that any escaping atmosphere extends beyond the Roche lobe radius ($R_{\rm Roche}$) of the planet. Therefore, a transit signature that is contemporaneous with the passage of the planet in front of the star (so the escaping atmosphere originates from the planet), yet indicates an obscuring area that places it outside the planet's Roche lobe (so the atmosphere is no longer bound to the planet) is a demonstration of atmospheric escape in action. Such a requirement indicates that the fractional dimming of the star, $\delta$, must exceed:
\begin{equation}
\delta \ge \left(\frac{R_{\rm Roche}}{R_*}\right)^2 \approx 0.13 \left(\frac{a}{0.025{\rm~AU}}\right)^2 \left(\frac{M_p}{M_J}\right)^{2/3}\left(\frac{R_*}{R_\odot}\right)^{-2}\left(\frac{M_*}{M_\odot}\right)^{-2/3}
\end{equation}
with $R_*$ and $M_*$ the stellar radius and mass, in units of the Sun's radius ($R_\odot$) and mass (M$_\odot$) respectively, $a$ is the planet's orbital separations, $M_p$ the planet's mass in units of Jupiter's mass $M_J$. An approximately $10$\% dimming far exceeds that possible from broadband transit measurements, which are typically an order of magnitude smaller for a hot Jupiter. However, line-absorption in planetary outflows can yield high optical depths and thus a large obscuration. 

The most commonly used method is a transit measurement in the Lyman-$\alpha$ line. The Lyman-$\alpha$ cross-section at line centre is $\sigma_{\rm \alpha,c}=5.9\times10^{-14}$~cm$^{2}$ for $10^4$~K gas \citep[e.g.][]{Hansen2006}.  For an obscuring region with an optical depth of order unity, with a scale size of order the planet's radius ($R_p$) the number density of neutral Hydrogen($n_{HI}$) at this point is very low:
\begin{equation}
n_{HI}\approx \frac{1}{\sigma_{\alpha,\rm c}R_p}= 2.4\times10^{2}\,{\rm cm}^{-2}\left(\frac{R_p}{R_J}\right)^{-1}
\end{equation}
where $R_J$ is Jupiter's radius. Therefore, the presence of small quantities of neutral Hydrogen gas at large distances from the planet can be readily detected through a Lyman-$\alpha$ transit. The Lyman-$\alpha$ transit was detected for the first transiting planet HD209458b shortly after its discovery \citep{Vidal-Madjar2003} with an absorption depth of $\sim15$\%, compared to the broadband, visible absorption depth of $\sim$1.5\% \citep{Charbonneau2000}. This transit depth indicated neutral Hydrogen is extended to large radii and represented the first good observational evidence of an exoplanet undergoing atmospheric escape. 

Subsequent Lyman-$\alpha$ transits were detected for the hot Jupiter HD189733b \citep{Levavelier2010} with an absorption depth of $5$\%.  A spectacular Lyman-$\alpha$ transit was detected for the close-in Neptune mass planet GJ436b \citep{Kulow2014,Ehrenreich2015}, with an absorption depth of 56\% compared to the broadband optical absorption depth of 0.69\% \citep{Ehrenreich2015}. There is no doubt that GJ436b is undergoing atmospheric escape, and this result was confirmed by \citet{Lavie2017} who observed the Lyman-$\alpha$ transit over a wide range of phases. 

\subsection{The details of Lyman-$\alpha$ transit spectroscopy}
\begin{figure}
\centering
\includegraphics[width=0.75\columnwidth]{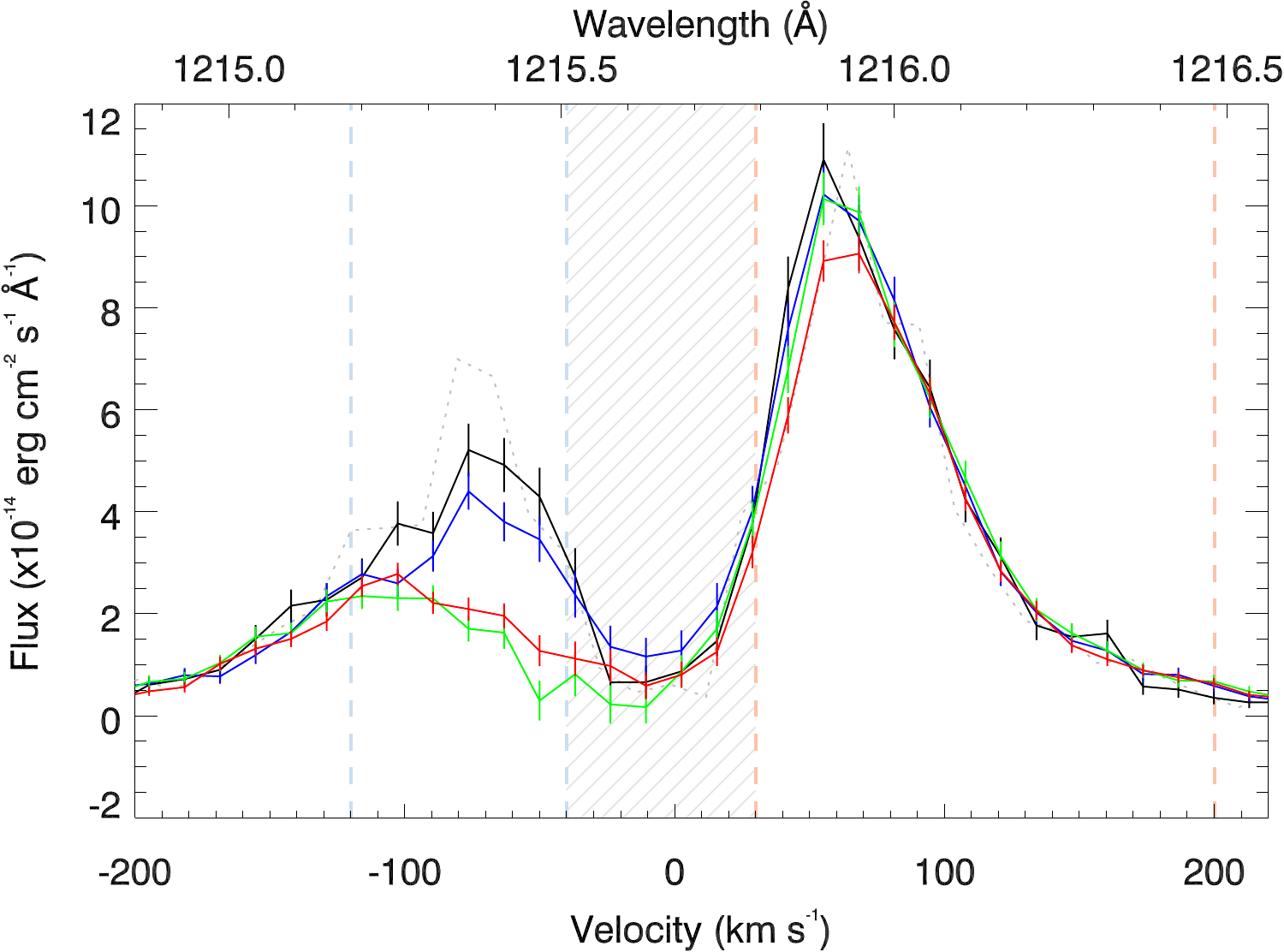}
\caption{The observed Lyman-$\alpha$ transit transmission spectra of GJ436. The solid lines show the averaged spectra taken over several visits. The black line shows the average out of transit spectra; the blue line shows the average pre-transit spectra; the green line shows the average in-transit spectra and the red line shows the average post-transit spectra. The dotted grey line shows a single out of transit observation. The hatched grey region indicates those parts of the spectra that are strongly affected by interstellar medium absorption (Gas between the star and Earth absorbs the emitted Lyman-$\alpha$ photons before they reach Earth). Note the large $\sim 50\%$ in-transit and post-transit reduction in measured Lyman-$\alpha$ flux at blue-shifted velocities ($\sim -80$~km~s$^{-1}$), indicating neutral Hydrogen in the planetary outflow is absorbing the Lyman-$\alpha$ photons emitted by the star; the pre/post-transit asymmetry indicates the escaping neutral Hydrogen is asymmetric and shaped into a cometary tail. Figure from \citet{Ehrenreich2015}. }\label{fig:GJ436b_Lya_transit}
\end{figure}

While a significant absorption in a Lyman-$\alpha$ transit is an excellent probe of atmospheric escape, the details of the transit spectrum are still uncertain. Figure~\ref{fig:GJ436b_Lya_transit} shows the spectrum of the Lyman-$\alpha$ line from GJ436 at several different phases of GJ436b's orbit. 

Absorption of the Lyman-$\alpha$ line by neutral Hydrogen gas in the interstellar medium and geo-coronal emission renders the core of the Lyman-$\alpha$ line useless for interpretation \citep[e.g.][]{Vidal-Madjar2003} (the hatched grey region in Figure~\ref{fig:GJ436b_Lya_transit}). The Lyman-$\alpha$ spectrum of GJ436 shows a deep absorption at blue-shifted velocities of $\sim -80$~km~s$^{-1}$ and a small absorption on the red-shifted side ($\sim 70$~km~s$^{-1}$). A deep absorption at blue-shifted velocities of order $100$~km~s$^{-1}$ is a typical signature in the Lyman-$\alpha$ transit of atmospheric escape. Such a velocity is puzzling from the concept of pure hydrodynamic escape. Hydrodynamic escape typically results in gas velocities which are of order the sound speed, which for $10^4$~K gas is $\sim 10$~km~s$^{-1}$ \citep[e.g.][]{Yelle2004,MurrayClay2009}, an order of magnitude smaller than those velocities observed. Thus, some other process is involved in accelerating neutral Hydrogen atoms to higher velocities. 

Two suggestions for the high velocities have been proposed in the literature. Radiation pressure can accelerate the gas to velocities $\sim 100$~km~s$^{-1}$ relative to the star and has been studied with particle simulations that do not include the hydrodynamics of the outflow \citep{Bourrier2013,Bourrier2014,Ehrenreich2015,Beth2016}. While these simulations are generally successful in reproducing the $\sim 100$~km~s$^{-1}$ blue-shifts observed, they struggle to reproduce the very high velocity absorption seen in some of the Lyman-$\alpha$ transits \citep{Bourrier2013}.

An alternative origin for the Lyman-$\alpha$ absorption at high velocities is charge exchange with solar-wind protons, resulting in hot, rapidly moving neutral Hydrogen atoms. These so-called Energetic Neutral Atoms (ENAs) are observed in the interactions between the solar wind and planetary atmospheres in the solar-system \citep[e.g.][]{Galli2008}. \citet{Holmstrom2008} suggested that the high-velocities observed in Lyman-$\alpha$ transits could be explained by the generation of ENAs at the interface between the planetary outflow and the stellar wind. \citet{Tremblin2013} used hydrodynamic simulations to study this interaction in multi-dimensions showing that ENAs were consistent with the observations. It is, of course, likely that both ENAs and radiation pressure are responsible for generating the high velocities observed in Lyman-$\alpha$ transits. In fact the modelling of the GJ~436b's transit that includes both radiation pressure and interactions with the stellar wind \citep{Bourrier2015,Bourrier2016} agrees very well with the observations \citep{Lavie2017}. Modelling that includes both radiation-pressure and ENAs within a hydrodynamic framework is required to explore this in more detail for all types of exoplanets.

\subsection{Other observational probes of atmospheric escape}

The Lyman-$\alpha$ line, due to its large cross-section, makes it the best probe to prove atmospheric escape is occurring, but there are several other indicators of atmospheric escape that are worth discussing. 

UV transit spectroscopy of HD209458b revealed large transit depths in lines of OI and CII \citep{Vidal-Madhar2004} and similarly for HD189733b \citep{BenJaffel2013}. These observations of heavy elements in the extended atmosphere are essential, as they reveal atmospheric escape must be occurring through a hydrodynamic process. Further work using FUV/UV spectroscopy also provides further constraints on the outflows, presents evidence for planet-star interactions and constrains
the high-energy output of the star \citep[e.g.][]{Fossati2015,Fossati2018}.  This conclusion is further backed up by the large transit radii detected using X-ray transits. X-ray photons are absorbed by heavy elements \citep[e.g.][]{Verner1995} and an absorption depth of $\sim 5-8\%$ has been detected in a broadband X-ray transit for HD189733b \citep{Poppenhaeger2013}, again confirming the presence of heavy-elements in the extended escaping atmosphere of hot Jupiters. 

Finally, recent observations of an extended He transit around WASP 107b has opened up a new probe of atmospheric escape \citep{Spake2018}. The 10833~\AA~ absorption feature of excited metastable Helium was predicted as a probe of the UV heated upper layers of an exoplanet atmosphere by \citet{Seager2000}, and its importance in the context of probing atmospheric escape was realised by \citet{OH2018}. EUV photons can ionize Helium. When a single ionized Helium atom recombines, approximately three-quarters of the recombinations populate the triplet levels, whereas the remaining quarter populate the singlet levels \citep[e.g.][]{Osterbrock2006}. The recombinations to the triplet states eventually lead (through radiative transitions) to the 2$^3$S level, which is  only able to radiatively decay through a forbidden-line photon, with a transition probability of $10.9$~day$^{-1}$ \citep[e.g.][]{Drake1971}, a timescale comparable to the flow timescale. Thus, absorption of 10833~\AA~photons (which excite the Helium atom to the 2$^3$P level) yield a significant transit signal \citep{OH2018}, with \citet{Spake2018} finding an absorption of $4.9\%$ for WASP 107b. The population of the 2$^3$S level of Helium is sensitive to the UV spectrum of the star \citep{OH2018}, meaning not all planets undergoing atmospheric escape are amenable to this probe \citep{Kreidberg2018}. However, the unique property of this probe is it can in principle be measured with ground-based telescopes rather than space-based. Accessibility from the ground means that high-resolution spectroscopy (with resolving powers of up-to $R\sim 100,000$) is achievable,  opening up the possibility of studying the kinematics of the outflow as it is launched from the planet.

\section{THE DYNAMICS OF ATMOSPHERIC ESCAPE}\label{sec:theory}
In this section, we shall focus entirely on atmospheric escape arising from heating of the planet's atmosphere (often called thermal escape). While non-thermal processes, often involving the interaction of the atmosphere with charged particles, can be significant, especially in Solar-System bodies today \citet{Lammer2008,Tian2015}, the extreme heating that close-in planets receive implies thermal escape processes dominate. 

One of the most important controlling parameters in atmospheric escape is the ratio of the gravitational binding energy of a gas particle to its thermal energy:
\begin{equation}
\lambda = \frac{GM_p\mu m_H}{k_bTR_p}\equiv\frac{GM_p}{c_s^2R_p}\approx 2.3\,\mu \left(\frac{M_p}{M_J}\right)\left(\frac{R_p}{R_J}\right)^{-1}\left(\frac{T}{10^4\,{\rm K}}\right)^{-1}
\end{equation}
where $G$ is the gravitational constant, $k_b$ is Boltzmann's constant, $\mu$ is the mean particle mass in units of the mass of a Hydrogen atom ($m_H$) and $c_s$ is the isothermal sound speed for gas at a temperature $T$. It is important to emphasise that it is not the equilibrium temperature we should be considering here, but the temperature of the upper atmosphere that is heated by UV and X-ray photons, where the typical temperatures are much higher \citep[e.g.][]{Gross1972,Lammer2003}. In the case that $\lambda < 1.5$, the internal energy of the gas exceeds the gravitational binding energy. In this case, the atmospheric gas simply flows out of the planet's atmosphere \citep{Opik1963}, sometimes referred to as ``blow-off''. 

\subsection{Jeans Escape}

When $\lambda$ is large, the gas particles are tightly bound to the planet, and atmospheric escape occurs through a process known as Jeans escape \citep{Jeans1925}. 
At higher altitudes in the atmosphere, the gas density is lower. As the gas density drops, so does the frequency of collisions between gas particles. Eventually, the gas becomes so rarefied that the average distance an individual gas particle moves before colliding with another (the mean-free-path) is larger than the scale height of the atmosphere itself and is termed the ``exobase''. The ratio of the mean free path to the representative scale length of the fluid ($\ell$) is known as the Knudsen number ${\rm Kn}$. Gas which has ${\rm Kn} \ll 1$ obeys the equations of fluid mechanics, while for gas which has ${\rm Kn\gtrsim 1}$ the equations of fluid mechanics are no longer appropriate. In Jeans escape, gas particles which exceed the escape velocity at the exobase are likely to escape, whereas those particles with velocities below the escape velocity cannot escape. Since $\lambda$ can be equivalently written as $\lambda\equiv 3/2(v_{\rm esc}/v_{\rm rms})^2$, then for $\lambda\gg 1$ the average gas particle velocity ($v_{\rm rms}$) is much smaller than the escape velocity. Thus, only a small fraction of the gas particles present at the exobase can escape. \citet{Tian2015} reviewed Jeans escape in the context of exoplanets, and we do not repeat it here. \citet{Lecavelier2004}, studied the Jeans escape in the context of hot Jupiters showing that unless the exobase extended all the way to the Roche lobe radius (where gas was simply able to free stream away from the planet, a process they called ``geometrical blow-off''), Jeans escape rates were small and unlikely to have an evolutionary impact on hot Jupiters.

\subsection{The hydrodynamics of escape}
In the context of highly irradiated exoplanets with large-scale heights, it is not {\it a priori} obvious that the exobase must occur in a region where the gas is still bound to the planet (i.e. the value of $\lambda$ that one would infer for a hydrostatic atmosphere at the exobase is $\gtrsim 1$). In such a scenario the atmospheric scale height $\rightarrow \infty$, and the gas density (and pressure) at large radius become constant. Such a scenario is exactly the same as that considered by \citet{Parker1958} in the context of a hydrostatic solar corona. Parker's insight was that without a large external pressure to balance the constant pressure of a static corona a hydrostatic solution is not in equilibrium, only a hydrodynamic outflow could produce a falling density and pressure at large distances. \citet{Parker1958,Parker1960} argued that there was only one possible hydrodynamic solution: a flow that started off sub-sonic, transitioned through a critical point where the gas velocity was equal to the sound speed (called the sonic point) and finally expanded supersonically with the density falling approximately as $1/r^2$ at large radius \citep[e.g.][]{Lamers1999}. This ``transonic'' solution is unique, and the position of the sonic point ($r_s$) is given by the solution to \citep{Parker1964}:
\begin{equation}
c_s^2= \frac{\partial \psi_g}{\partial\log A}{\rm ;}\quad\Rightarrow r_s\approx\frac{GM_p}{2c_s^2}\approx 6.4~R_p \left(\frac{M_p}{{\rm M}_J}\right)\left(\frac{c_s}{10\,{\rm km~s}^{-1}}\right)^{-2}\left(\frac{R_p}{1.5\,{\rm R}_J}\right)^{-1}\label{eqn:sonic}
\end{equation}
for an isothermal outflow, where $\psi_g$ is the potential from which the outflow is escaping, and $A$ is the stream-bundle area. The further approximate equalities are for a spherical outflow from a point mass potential. The role of the sonic point is critical. As the flow beyond the sonic point is travelling super-sonically, it is unable to propagate any information upstream. Therefore, should the flow become collisionless (such that the equations of hydrodynamics no longer apply) at a radius larger than the sonic point this cannot affect the outflow from the planet. Furthermore, if the outflow were to interact with the stellar wind/magnetic field outside its sonic-point, this can have no impact on the outflow emanating from the planet. Alternatively, if the flow becomes collisionless before the sonic point, the transonic solution is not correct. It is gas pressure gradients that accelerate the gas towards sonic velocities; these cannot operate in a collisionless fluid, so the flow can never reach super-sonic velocities. Instead, the gas density/pressure profile is closer to hydrostatic and mass-loss is likely to proceed through Jeans escape, where the velocity distribution at the exobase is slightly shifted with an outward velocity profile \citep{Yelle2004,Volkov2011a,Volkov2011b}.

\subsection{``Energy-limited'' mass-loss} \label{sec:EL}

Before we go onto discuss the developments in the numerical simulations of hydrodynamic escape, it is useful to review what is known as ``energy-limited'' escape. It may seem asinine to discuss an energy-limit, as surely basic physical conservation laws mean the outflows should be bound by the conservation of energy. However, the term has come to mean a specific form of a mass-loss prescription, which has evolved slightly from the original description of energy-limited outflows. \citet{Watson1981}'s calculations of hydrodynamic escape from Earth and Venus forms the basis for energy-limited mass-loss. The original description is as follows: a planet absorbs all the high-energy flux in a thin layer at a radius $R_{\rm HE}$ around the point where the optical depth to high-energy photons is unity. In the absence of other heating or cooling sources, the supplied high-energy flux ($F_{\rm HE}$) must balance the work done escaping from the planet's potential. Thus, the mass-flux per steradian is:
\begin{equation}
\dot{M} = \frac{F_{\rm HE}R_{\rm HE}^2R_p}{GM_p}
\end{equation}
The calculation of the mass-loss rate then reduces to the calculation of $R_{\rm HE}$. The naive guess might be that as you increase the gas density in the atmosphere, the high-energy photons are absorbed at larger distances from the planet, and the mass-flux increases. However, this is not necessarily true. Since there is no atmospheric heating source between $R_p$ and $R_{\rm HE}$, as gas moves away from the planet it expands and cools. \citet{Watson1981} argued conduction offsets this cooling. Ultimately, the maximum mass-loss rate is obtained when the gas temperature between $R_p$ and $R_{\rm HE}$ approaches zero. Any increase in atmospheric density pushes $R_{\rm HE}$ to a larger radius, reducing the temperature gradient and hence the conductive heat flux towards $R_p$ as the gas temperature cannot fall below zero. This reduction in conductive heat flux must be counter-balanced by a reduction in $P{\rm d}V$ cooling by lowering the mass-flux. Therefore, in \citet{Watson1981}'s parlance, the mass-flux is ``energy-limited'' by the upstream conductive heat-flux to balance the adiabatic cooling of the expanding gas as it approaches $R_{\rm HE}$. \citet{Watson1981} presented a formalism to calculate the maximal $R_{\rm HE}$ based on this criterion. \citet{Lammer2003} used \citet{Watson1981}'s formalism to derive hydrodynamic escape rates from hot Jupiters.
These calculations gave much larger escape rates up to $10^{12}$g~s$^{-1}$, than previous work that had tended to use estimates from Jeans escape, which gave rates $\ll 10^{10}$~g~s$^{-1}$.

The use of an energy-limited mass-loss expression is common in the literature; however, the connection to the original ``energy-limit'' of downward conduction of heat has been lost. The total mass-loss rate is often written as:
\begin{equation}
\dot{M} = \eta \frac{\pi R_p^3F_{\rm HE}}{GM_p K_{\rm eff}} \label{eqn:EL_main}
\end{equation}
The term $K_{\rm eff}$ was introduced by \citet{Erkaev2007} to account for the fact that the flow does not need to escape to infinity, but merely out of the planet's Roche lobe. This fact means the mass-loss rates can be larger. This inference can also be seen from Equation~\ref{eqn:sonic} where the weaker gradient of the planet and star's effective potential moves the sonic point closer to the planet to regions of higher density. Equation~\ref{eqn:EL_main} exists in the literature with many different pre-factors \citep[e.g.][]{Lecavelier2007,Erkaev2007,Jackson2012,Lopez2012,Kurokawa2014,Chen2016}. This difference arises since planets absorb high energy flux over an area $\pi R_{\rm HE}^2$, but have surface areas of $4\pi R_{\rm HE}^2$. Without detailed knowledge of the energy redistribution one has to choose how to average this incoming energy over the planet's surface. Such choices are moot, as Equation~\ref{eqn:EL_main} contains an efficiency parameter ($\eta$) that is included to incorporate the (lack of) knowledge of energy-gain and loss processes\footnote{Due to the different pre-factors in use, care must be taken when comparing one efficiency value to another}. Equation~\ref{eqn:EL_main} presents several problems for calculating the mass-loss rates: (i) It is unclear what wavelength range high-energy flux needs to be incorporated in $F_{\rm HE}$; stars emit high-energy photons non-thermally over a wide range of wavelengths from the Far-UV to the hard X-rays. (ii) It is unclear what value of $\eta$ to choose and whether to assume it is constant. (iii) Finally, Equation~\ref{eqn:EL_main} presents no way to determine whether hydrodynamic escape is appropriate, rather than Jeans escape which would yield a much lower mass-loss rate. It is for these reasons that we must appeal to numerical calculations to determine the mass-loss rates and to determine whether hydrodynamic escape is appropriate.

\subsection{Numerical models of hydrodynamic escape}
There are currently no complete numerical simulations of hydrodynamic escape that include all the necessary radiative transfer, thermodynamics and chemistry even in one-dimensional calculations. Therefore, approximations have been made to one or several of the different important ingredients. Apart from the few multi-dimensional calculations, all one-dimensional calculations start from the same basic setup. Close-in exoplanets are expected to be tidally locked, such that they have a permanent day-side and night-side. Comparing the ratio of the gravitational acceleration ($g$) to the acceleration arising from the Coriolis force ($a_c$) one finds:
\begin{equation}
\frac{g}{a_c}\sim \frac{g}{c_s\Omega}\sim\left(\frac{a}{H}\right)\left(\frac{c_s}{v_K}\right)\approx 0.1 \frac{a}{H}
\end{equation}
where $\Omega$ is the planetary orbits angular frequency and $H$ the scale height. Since $H\lesssim R_p$ below the sonic point and $a/R_p\gg10$ for the observed exoplanets, the gravitational force dominates over the Coriolis force inside the sonic point \citep{MurrayClay2009}. This result means the streamlines, in the sub-sonic regime, will not be strongly bent by the Coriolis force. Thus a streamline that leaves the sub-stellar point on the planet will roughly follow the locus connecting the star and planet. Therefore, the basic setup of most one-dimensional calculations is to assume a streamline with a spherical divergence that connects the star and planet. One then proceeds to solve the radiative-transfer and hydrodynamic problem along this streamline. 

\citet{Yelle2004} studied atmospheric escape from hot Jupiters, using the properties of HD209458b as a reference. In this calculation, a Hydrogen/Helium ionization and a chemical network were employed including conduction; also \citet{Yelle2004} assumed that 63\% of the photoelectrons energy was converted into heat based on simulations of Jupiter's atmosphere \citep{Waite1983}. \citet{Yelle2004}'s simulations showed that the mass-loss rates typically scaled linearly with input energy flux, that at the fluxes experienced by old hot Jupiters $P{\rm d}V$ work dominated the cooling and that conduction was of limited importance.  \citet{Yelle2004} also demonstrated that the energy-limited calculations of \citet{Lammer2003} typically overestimated the mass-loss rates by a factor of roughly 20 and that the mass-loss rates from hot Jupiters were too small to remove significant amounts of mass over their lifetime. The overestimation by \citet{Lammer2003}'s calculation primarily arises from the fact the \citet{Watson1981} model does not include radiative cooling. 

\citet{Tian2005a} studied a similar problem to \citet{Yelle2004}, but instead of calculating the production of photoelectrons directly, a constant fraction of the absorbed radiative energy was deposited into heat, where \citet{Tian2005a} varied the values between 10\% and 60\%. These calculations again indicated that applying \citet{Watson1981}'s approach to hot Jupiters significantly overestimated the mass-loss rates. \citet{Tian2005a} also checked the validity of the hydrodynamic approximation demonstrating that the exobase occurred downstream of the sonic point indicating that atmospheric escape from close-in exoplanets was indeed transonic and hydrodynamic. \cite{Yelle2004} and \citet{Tian2005a}'s approach of a constant heating efficiency has been adopted by a wide number of further calculations studying both hot-Jupiters \citep{Munoz2007} and super-Earth/mini-Neptune mass planets \citep{Lammer2013,Lammer2014,Lammer2016,Erkaev2013,Erkaev2016}. The key results of these works are that atmospheric escape for giant planets is unlikely to have an evolutionary impact, while it will be significant for close-in low-mass planets with Hydrogen-rich atmospheres. 
Recently, this style of calculations has been improved where the adopted constant local heating efficiencies are determined from Monte-Carlo calculations of the deposition of a photoelectron's energy \citep{Shematovich2014,Ionov2015,Ionov2017,Ionov2018}, indicating the heating efficiency never exceeds $\sim 20\%$. 

\citet{MurrayClay2009} studied atmospheric escape from hot Jupiters but performed a different style of calculation. By studying a coupled radiative-transfer and hydrodynamic calculation of a pure Hydrogen atmosphere, \citet{MurrayClay2009} showed there were distinctly two regimes of  hydrodynamic escape from hot Jupiters, a regime at high-fluxes where the mass-loss rate scaled approximately with the square-root of the UV flux, and a regime at low-fluxes where the mass-loss rate scaled linearly with input UV flux. At high-fluxes, radiative recombinations of Hydrogen produced enough ionizing photons to affect the ionization of Hydrogen deep in the atmosphere. This two-body process scales as density squared; balancing incoming ionizing photons with recombinations, one finds the density of ionized Hydrogen scales with the square root of the incoming ionizing flux \citep{Osterbrock2006}. At low-fluxes, the lower densities resulted in much longer recombination times. As recombination is one of the dominant cooling mechanisms the energy-lost to radiative cooling was small \citep{MurrayClay2009} and $P{\rm d}V$ work dominated the gas cooling. In this case, the flow became closer to the energy-limited case and scaled linearly with incoming UV flux. \citet{Bear2011} and \citet{OA2016} demonstrated this transition from energy-limited to recombination-limited UV driven outflows applied to a wide variety of planets, not just hot Jupiters, where the limiting criterion was that recombination-limited flows occurred when the recombination timescale was shorter than the flow timescale, and the energy-limited case occurred when the recombination timescale was longer than the flow timescale. 

\citet{Owen2012} used results from detailed X-ray and UV radiative transfer calculations to determine the gas temperature as a function of the ionization parameter ($4\pi F_{\rm HE}/n$, with $n$ the number density of the gas). This temperature-ionization parameter relation was then used to carry out hydrodynamic calculations over a wide range of parameter space. \citet{Owen2012} showed that the ``efficiency parameter'' was far from constant, generally decreasing with increasing depth of the gravitational potential, due to more energy loss from radiative cooling with longer flow timescales. Furthermore, this work demonstrated that typically efficiencies for super-Earths and mini-Neptunes was $\sim 10$\% and that the X-rays generally drove the flow at high-fluxes and the UV at low-fluxes.

\subsubsection{Multidimensional calculations}\label{sec:multi-d}

\begin{figure}
\centering
\includegraphics[width=0.63\textwidth,trim={0cm 30.3cm 0cm 0cm},clip]{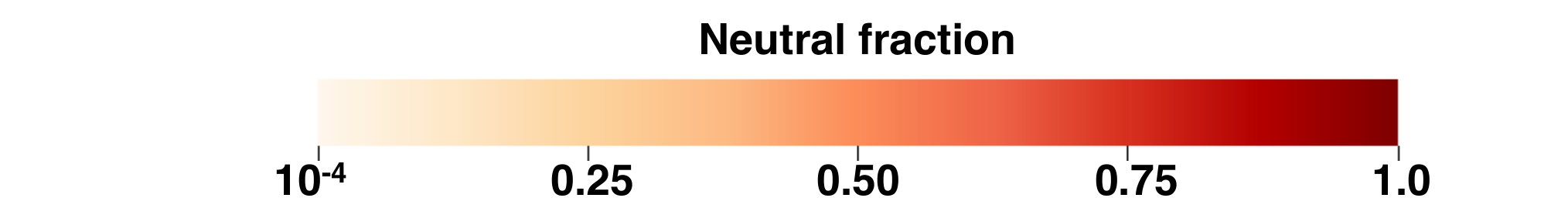}
\includegraphics[width=0.63\textwidth,trim={0cm 0cm 0cm 16.98cm},clip]{f6}
\caption{The velocity (vectors) and ionization structure (colormap) of an outflow from a hot Jupiter with a radius of 2.14~R$_J$ and mass of 0.53~M$_J$. The star is located to the left of the computational grid. Note the flow from the day-side to the night-side at the terminator and the Kelvin-Helmholtz like rolls. Figure from \citet{Tripathi2015}.} \label{fig:3dflow}
\end{figure}

There have been a few multi-dimensional calculations of the launch of hydrodynamic outflows. \citet{Stone2009} showed that a planet's permanent day- and night-side resulted in anisotropic heating of the upper atmosphere. In their two-dimensional simulations, they deposited heat on the day-side of the planet's atmosphere. The large pressure gradient that occurred at the terminator of the planet caused the flow to wrap around to the night-side as shown in Figure~\ref{fig:3dflow}. This picture of rapid day-to-night side flow around the terminator was further confirmed by \citet{Owen2014} who incorporated a simple EUV radiative transfer scheme into a two-dimensional hydrodynamic calculation and \citet{Tripathi2015} who employed the method of \citet{MurrayClay2009} in a three-dimensional simulation; the resulting flow topology found by \citet{Tripathi2015} is shown in Figure~\ref{fig:3dflow}. Furthermore, the three-dimensional calculations indicate that the flow on the night-side is unsteady, possibly coming from Kelvin-Helmholtz instabilities that arise from the rapid shear present in the day-to-night flow, but that the average mass-loss rates were similar to those derived from one-dimensional calculations. 

Recent work has also looked at the dynamics on the largest scales and its interaction with the solar wind. In these calculations, the launch from the planet is ignored, the mass-loss rates are parametrised, and the outflow is often taken to be a Parker wind \citep{Schneiter2007,Schneiter2016,Bisikalo2013,CarrollNellenback2017,Debrecht2018}. 

\subsection{The role of magnetic fields}

One of the least explored aspects of atmospheric escape from close-in exoplanets is the importance of magnetic fields. The extreme high-energy irradiation of the outflows means it is highly ionized (e.g. Figure~\ref{fig:3dflow}) and hence coupled to any planetary magnetic field \citep{Adams2011,Trammell2011}.

\citet{Adams2011} and \citet{Trammell2011} argued that a strong dipole magnetic field would result in closed field lines in the vicinity of the sub-stellar point, but that at higher latitudes the field lines would be opened out by the thermal pressure of the UV/X-ray heated atmosphere, or any stellar magnetic field, resulting in outflow. This picture was confirmed by two-dimensional numerical simulations that assumed an isothermal outflow \citep{Trammell2014} or included radiative transfer \citep{Owen2014}. Allowing outflow only along open field lines at latitudes above the sub-stellar point significantly reduced the mass-loss rates below those calculations without strong magnetic fields, or those assumed from one-dimensional calculations that solve for the flow solution only from the sub-stellar point. For field strengths of a few Gauss (comparable to Jupiter's magnetic field) around hot Jupiters, \citet{Owen2014}, \citet{Khodachenko2015} and \citet{Arakcheev2017} found the mass-loss rates were suppressed by an order of magnitude. The effect of planetary magnetic fields around lower mass super-Earths/mini-Neptunes remains unexplored; however, an order of magnitude correction to the mass-loss rates typically taken from one-dimensional calculations would have significant implications on our understanding of the evolution of close-in, low-mass exoplanets. 

Like the pure hydrodynamic calculations, work has also been done on the interaction of the flow with the broader interplanetary environment, where the interaction between any planetary magnetic field, the outflow and the stellar magnetic field ultimately controls what happens to the gas that has escaped the planet \citep{Matsakos2015,Alexander2016,DaleyYates2017,Villarreal2018}.

\section{THE LONG-TERM EVOLUTION OF EXOPLANETS UNDERGOING ATMOSPHERIC ESCAPE}\label{sec:evolve}
The evolution of a close-in planet's Hydrogen/Helium atmosphere is governed by two processes: the cooling and contraction of the atmosphere and mass-loss arising from atmospheric escape. 

Planets accrete their primordial Hydrogen/Helium atmospheres from their parent protoplanetary discs.
A planet's atmosphere starts off with high entropy and is extended and then cools by contracting under gravity becoming denser and smaller.
This cooling means that planets are largest when they are young, and therefore able to absorb a more significant fraction of the star's high-energy luminosity resulting in higher mass-loss rates. 

\subsection{The evolution of a star's high-energy flux}
An additional effect compounds the larger planetary radii at young ages leading to higher mass-loss rates: stars emit a higher fraction of their total luminosity non-thermally at high energies when they are young. 
The evolution of a star's X-ray luminosity ($L_X$) as a function of time typically has the following pathway: at young ages the X-ray luminosity is roughly a constant fraction of the bolometric luminosity of the star and is considered ``saturated'' at a value $L_{\rm sat}$, \citep{Guedel2004}. This ``saturation timescale'' ($t_{\rm sat}$) typically lasts 100~Myr \citep{Jackson2012,Tu2015}. After this saturation period ends the stellar X-ray luminosity decays with time. 
The stellar X-ray luminosity typically falls faster than $1/{\rm age}$. The surface extreme ultra-violet (EUV) flux is proportional to the surface X-ray flux \citep{Chadney2015}, with higher X-ray fluxes resulting in a higher fraction of the star's high-energy luminosity being emitted in the X-rays rather than the ultra-violet (UV).   As such the high-energy ``exposure'' ($\int\! L_{\rm HE}\,{\rm d}t$) is dominated at early times rather than at late times. In fact, atmospheric escape is most important when the high-energy flux is saturated \citep{Lopez2013,Owen2013}. The evolution of a star's high-energy flux is typically parametrised as \citep[e.g.][]{Ribas2005,Jackson2012}:
\begin{equation}
\frac{L_{\rm HE}}{L_{\rm bol}} = 
\begin{cases}
L_{\rm sat}\quad{ \rm{for}\,t\le t_{\rm sat} }\\
L_{\rm sat}\left(\frac{t}{t_{\rm sat}}\right)^{-1-\alpha}\quad{\rm{for}\,t>t_{\rm sat}}
\end{cases}
\end{equation}
where $\alpha$ is some positive value typically around $\sim 0.5$ \citep{Jackson2012,Tu2015}. 
%


Finally, it is also important to discuss how the evolution of the high-energy flux varies with stellar mass. \citet{Jackson2012}, showed that for lower-mass stars the saturation time-scale was longer and the value of $L_X/L_{\rm bol}$ at which they were saturated was slightly higher. Combining this fact with the longer pre-main-sequence lifetimes of lower-mass stars, \citet{Lopez2016} argued that the ratio of high-energy exposure to bolometric exposure scaled as $M_*^{-3}$, meaning at fixed equilibrium temperature atmospheric escape is significantly stronger around lower-mass stars. 

\subsection{Atmospheric escape driven evolution of close-in exoplanets}\label{sec:evolve_theory}

With knowledge of the mass-loss rates arising from atmospheric escape and the evolution of a star's high energy flux, we are now in a position to understand atmospheric escape's effect on an exoplanet's evolution. 
\subsubsection{The evolution of close-in giant planets}\label{sec:evolve_giants}
Many of the early studies of an exoplanet's evolution including escape focused, unsurprisingly, on hot Jupiters. \citet{Baraffe2004,Baraffe2005} coupled \citet{Lammer2003}'s reformulation of the \citet{Watson1981} energy-limited model to an evolutionary calculation. As described above \citet{Lammer2003}'s model found very high mass-loss rates ($\sim 10^{12}$~g~s$^{-1}$), much higher than those found from hydrodynamic calculations ($\sim 10^{10}-10^{11}$~g~s$^{-1}$, e.g. \citealt{Owen2012}). These high mass-loss rates caused the giant planets to undergo an interesting evolutionary pathway. 

The timescale for the thermal evolution of a Hydrogen/Helium dominated atmosphere is the Kelvin-Helmholtz contraction timescale ($t_{\rm KH}$): the timescale for the atmosphere to radiate away its current gravitational binding energy at its current luminosity. For planets, the Kelvin-Helmholtz timescale closely tracks the planet's age. The timescale relevant for the evolutionary impact of atmospheric escape is the mass-loss timescale: the timescale to remove the current atmosphere at the current mass-loss rate. 

Now by construction, the planet's initial Kelvin-Helmholtz timescale has to be shorter than the initial mass-loss timescale. The Kelvin-Helmholtz timescale increases roughly linearly with time as the planet ages, but as the atmospheric mass drops due to escape the mass-loss timescale can remain constant or even decrease with time. When the mass-loss timescale becomes shorter than the Kelvin-Helmholtz timescale, the planet is no longer able to thermally cool by radiation, and the interior essentially remains at constant entropy. Self-gravitating, convective objects have a mass-radius relationship where the radius increases for decreasing mass at constant entropy \citep[e.g.][]{Kippenhahn2012}. If a giant planet enters the stage where the mass-loss timescale is considerably shorter than the Kelvin-Helmholtz timescale, then when it loses mass it adiabatically adjusts to a larger radius. In the energy-limited model as $\dot{M}\propto R_p^3/M_p$, this leads to more substantial mass-loss rates causing the planet to expand faster giving even larger mass-loss rates. This positive feedback loop causes the planet to undergo run-away mass-loss. Ultimately, if the planet contains a solid core, the planet's radius will start to decrease with decreasing mass when the core's gravity dominates over the self-gravity of the atmosphere, stabilising the mass-loss rate. \citet{Baraffe2005} suggested that this process might be the origin of the low-mass planets, where by hot Jupiters undergo runaway mass-loss rates and become hot Neptunes. \citet{Kurokawa2014} extended this further by noting that during the run-away mass-loss phase the hot Jupiter would likely fill its Roche lobe and undergo Roche-lobe overflow resulting in even more rapid mass-loss. 

More realistic simulations of atmospheric escape from hot Jupiters \citep[e.g.][]{Yelle2004,MurrayClay2009} suggest that the energy-limited models of \citet{Lammer2003} significantly overestimated the mass-loss rates and they were, in reality, a factor of 20-100 times lower. When the lower mass-loss rates derived from hydrodynamic simulations are included in evolutionary calculations of hot Jupiters, they are found to be stable against atmospheric escape, losing at most a few percent of their initial atmospheric inventories \citep{Hubbard2007,Owen2013,Jin2014}. This result also means that hot Jupiters and close-in low mass planets are not genetically related through atmospheric escape, in agreement with other statistical arguments \citep{Winn2017}. 

\subsubsection{The evolution of close-in low-mass planets}

While it has become apparent that atmospheric escape is not a bulk evolutionary driver of giant planets, much of the early theoretical work on runaway mass-loss is vital in the context of the evolution of close-in low-mass planets. 

This discovery of close-in low-mass planets that required voluminous Hydrogen/Helium dominated atmospheres to explain their observed mass and radius led to an interest in their evolution. The structure of these planets is thought to be a solid core with a mass of several to tens of Earth masses surrounded by a large Hydrogen/Helium atmosphere. \citet{Baraffe2006} studied the evolution of close-in planets, adopting the energy-limited model of \citet{Lammer2003}, but unlike their earlier work also considered models where the assumed mass-loss rates were lower. 
Following the discovery of the first close-in rocky exoplanet (CoRoT-7b, \citealt{Leger2009}), \citet{Jackson2010} and \citet{Valencia2010} studied its possible evolutionary histories, arguing that even though it contained a negligible Hydrogen/Helium atmosphere today, its short orbital period of $\sim 0.8$~days would have allowed it to have been born with a significant Hydrogen/Helium atmosphere and have atmospheric escape completely remove it. 

One of the most significant hints of atmospheric escape's role in sculpting close-in, low-mass exoplanets was the discovery of the {\it Kepler}-36 system \citep{Carter2012}. The {\it Kepler}-36 system contains two planets with very similar orbital separations (b at 0.115~AU and c at 0.128~AU), but they have dramatically different densities. The inner planet's composition is consistent with being completely solid, while the outer planet has a substantial Hydrogen/Helium atmosphere, making up roughly 10\% of the planet's mass \citep{Carter2012}. The only difference being the inner planet is less massive ($\sim 4.4$~M$_\oplus$ compared to $\sim 8.1$~M$_\oplus$, with $M_\oplus$ the mass of the Earth). \citet{Lopez2013} demonstrated that both planets could have been born with Hydrogen/Helium atmospheres making up approximately 22\% of the planets' initial mass. The lower core-mass of the inner planet meant its gravitational well was too weak to resist complete removal of its atmosphere, while the deeper potential well of the outer planet meant it was able to retain about half its initial atmosphere. 

A key insight of this early work is that because low-mass planets have significantly large radii at young ages (in some cases up to 10~R$_\oplus$, \citealt{Lopez2012,Lopez2013}) and the received high energy fluxes were much higher as well, the majority of the mass-loss typically occurred in the first few 100~Myr of a planet's lifetime \citep{Lopez2013,Owen2013}. 

\begin{figure}
\centering
\includegraphics[width=\textwidth]{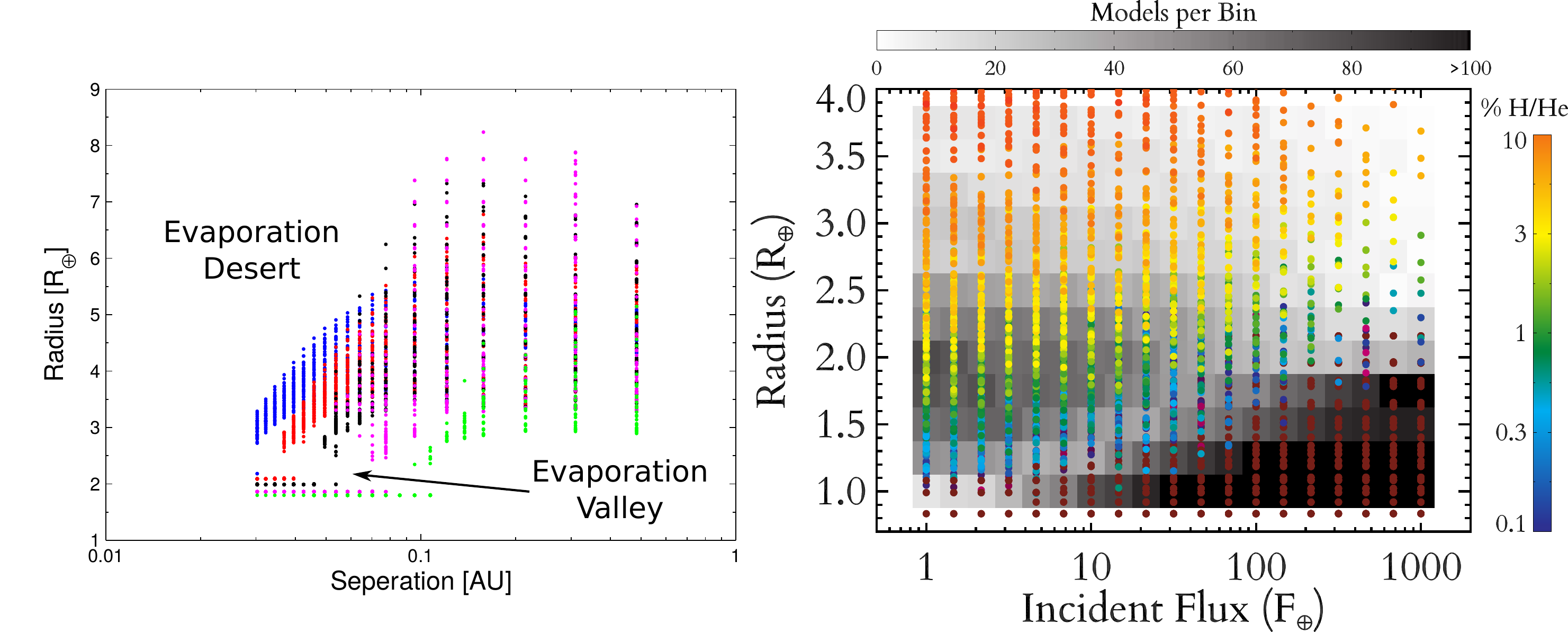}
\caption{Left panel: the radius-separation distribution of a 10~Gyr old exoplanet population that has experienced atmospheric escape during its evolution, with the planetary radii shown in units of the Earth's radius ($R_\oplus$). The planet population has core masses ranges between (15~M$_\oplus$, blue) and (6.5~M$_\oplus$, green) and initial total masses $<20$~M$_\oplus$. Figure taken from \citet{Owen2013}. Right panel: the occurrence rate in planet radius and incident bolometric flux in unit's of the Earth's ($F_\oplus$) (grey-scale) and atmosphere mass fractions (colours) of a 5~Gyr old exoplanet population that was initial uniformly distributed in log period, core mass and atmosphere mass fraction. Figure taken from \citet{Lopez2013}.  Note the presence of the evaporation desert and the fact that the evaporation valley has a slope where the planet radius is larger closer to the star.  }\label{fig:pop_evolve}
\end{figure}

With the growing number of small planets discovered, studies of a population of low-mass close-in exoplanets evolving under the influence of atmospheric escape became an area of interest. \citet{Owen2013} coupled the mass-loss rates computed by \citet{Owen2012} to evolutionary models of low-mass exoplanets. Again they adopted a structure of a solid core surrounded by a Hydrogen/Helium atmosphere. In this work, \citet{Owen2013} varied the core masses (from 6.5~M$_\oplus$ to 15~M$_\oplus$), orbital locations and initial atmospheric masses for a large synthetic planet population and followed its evolution under the influence of atmospheric cooling and escape. The resulting radius-separation distribution is shown in the left panel of Figure~\ref{fig:pop_evolve}. 

The resulting population of close-in exoplanets in the radius-period (or, equivalently, separation from the star) plane show several critical features of atmospheric escape driven evolution, for example, the lack of large planets on very short-orbital periods, later labelled the ``evaporation desert''. 
However, the most intriguing result was the other region of low planet occurrence: the empty band starting around 2~R$_\oplus$ at 0.03~AU and slowly declining in radius to 0.1~AU, later labelled ``the evaporation valley''. The models showed that those planets that resided below this region were planets that initially had Hydrogen/Helium atmospheres, but ultimately lost them due to atmospheric escape, while those just above this region had Hydrogen/Helium atmosphere masses of $\sim 1\%$ of the core mass. The fact that to completely strip a higher mass core required a higher flux resulted in the slope to this low-occurrence region (clearly seen in the right panel of Figure~{\ref{fig:pop_evolve}).  

\citet{Lopez2013} extended the study of \citet{Owen2013} to a broader range of core-masses but using an energy-limited prescription for mass-loss with a constant efficiency. The resulting radius-period distribution is shown in the right panel of Figure~\ref{fig:pop_evolve}. \citet{Lopez2013}'s population showed very similar features to the \citet{Owen2013} population, recovering the evaporation desert and the evaporation valley. While the exact boundaries of both the evaporation desert and valley were found to depend on the details of the atmospheric escape model, both features were identified as robust evidence of atmospheric escape driven evolution of close-in exoplanets. Further studies by \citet{Jin2014}, \citet{Howe2015} and \citet{Chen2016} found similar results. It is also important to emphasise the evaporation valley does not result from removing those planets that had atmospheres that initially placed them in the valley. Instead, planets with low enough core masses can start with very large atmospheres, cross the valley and be completely stripped \citep{Owen2017}. 

\citet{Owen2013} and \citet{Lopez2013} also made a valuable insight; since the evaporation valley provided an observable way to separate solid cores from those with volatile atmospheres it provided a way to constrain the densities and hence the compositions of the solid cores. The minimum separation at which a core of a particular mass could be stripped provides a constraint on its mass (provided the mass-loss rates are known from models), and the observed location of the valley provided a constraint on its radius. 

The origin of both the evaporation desert and valley can be understood simply by understanding the form of the atmospheric mass-loss timescale as a function of atmosphere mass \citep{Owen2017}. A schematic plot of this is shown in Figure~\ref{fig:tmdot}. 
\begin{figure}
\centering
\includegraphics[width=0.55\textwidth]{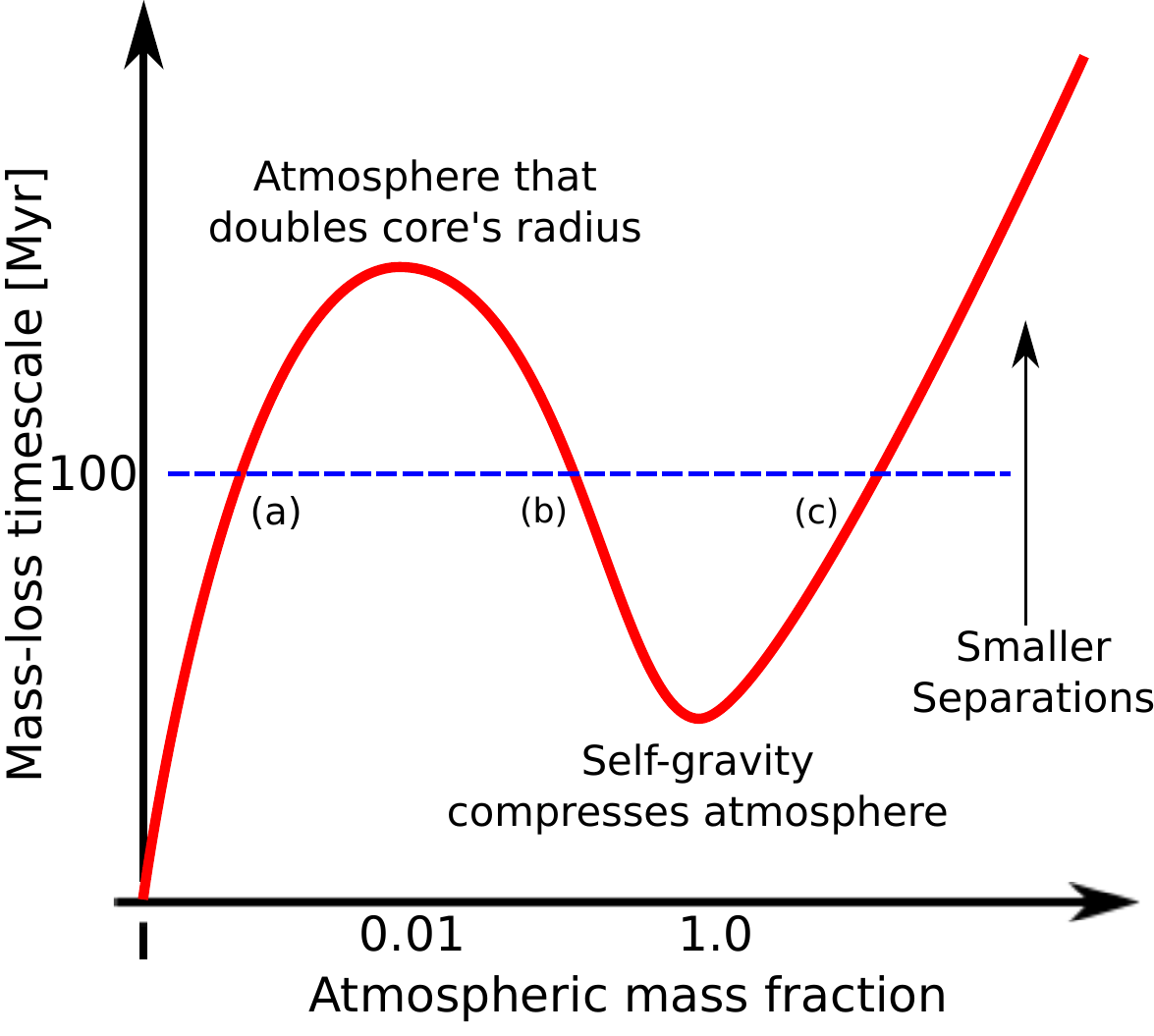}
\caption{Schematic plot showing the mass-loss timescale as a function of the atmospheric mass fraction for a planet with a fixed core mass. }\label{fig:tmdot}
\end{figure}
This diagram shows the mass-loss timescale has two turning points as the atmospheric mass fraction ($M_{\rm atm}/M_{\rm core}$) varies. The first turning point occurs when the atmospheric mass is large enough that the radius of the atmosphere is large enough to double the core's radius (typically containing $\sim 1\%$ of the core's mass). At this point, the atmosphere's radius dominates the radius of the planet, and its radius increases rapidly such that the extra absorbed high-energy flux counterbalances the increased atmosphere mass. This trend results in a decreasing mass-loss timescale with increasing atmosphere mass (b) until the atmosphere is so massive (such that it contains a comparable mass to the solid core) that self-gravity of the atmosphere compresses it resulting in a roughly fixed radius. At this point, the mass-loss time-scale begins to increase again (c). \citet{Owen2017} demonstrated it is this shape of the mass-loss timescale that produces the evaporation desert and valley. The blue dashed line in the diagram represents a fixed timescale (roughly $\sim 100$~Myr as discussed above) for mass-loss to occur; it moves up in the diagram as the planet moves to smaller separations or the core mass becomes lower. Any planet with an atmosphere mass that places it below this line is unstable to atmospheric mass-loss and will lose mass until it is completely stripped, or reaches an atmospheric mass that is stable to mass-loss. Therefore the ``evaporation valley'', which occurs between stripped cores and those that retain small envelopes is created by the complete removal of a planet's atmosphere if initially resides with an atmospheric mass fraction below point (a). The desert is created by planets with atmospheric masses that initially placed them between points (b) and (c) in the diagram, but evolve to lower mass (point b). 

\begin{figure}
\centering
\includegraphics[width=0.75\textwidth]{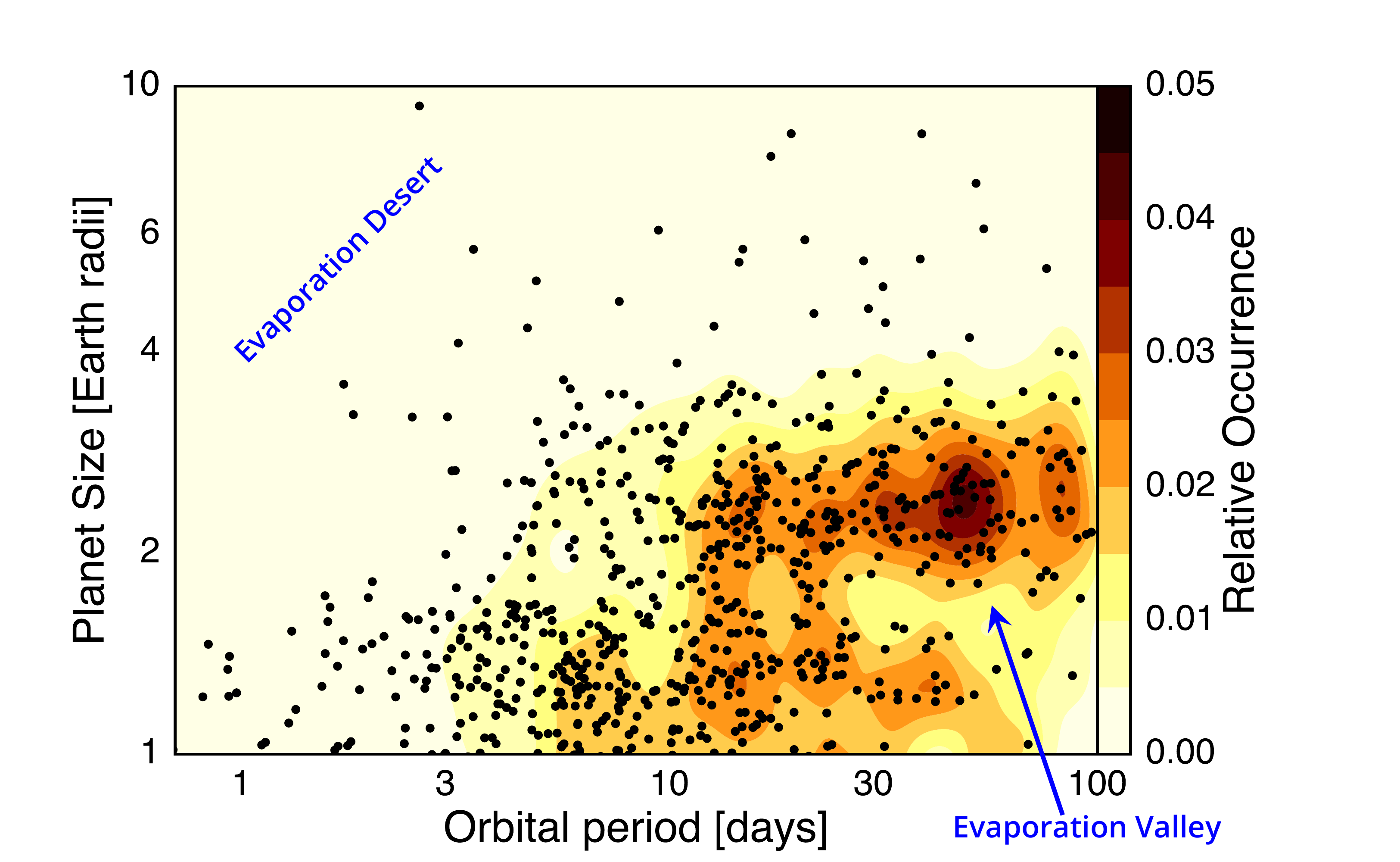}
\caption{The observational bias corrected planetary occurrence rate in the planetary radius and period planet as the colour map. The points represent planets studied by the California-Kepler-Survey (CKS). Note the evaporation desert and valley are clearly present, from \citep{Fulton2017}. }\label{fig:cks}
\end{figure}

\subsection{Signatures of atmospheric escape driven evolution of the exoplanet population}

Observations of atmospheric escape actually occurring in a handful of nearby highly irradiated exoplanets are useful in providing benchmark cases for our theoretical models. 
However, the majority of the observed exoplanets are billions of years old \citep[e.g.][]{Melo2006}, indicating that atmospheric escape we can observe today is not indicative of the powerful outflows that occurred early in a planet's life. 
Therefore, by looking at the properties of the old exoplanet population today, we can look for the imprints of atmospheric escape. 

By looking at the mass-period distribution of close-in exoplanets \citet{Szabo2011} identified a ``sub-Jovian'' desert where those planets with intermediate masses ($\sim 20-200$\,M$_\oplus$) were missing, out to orbital periods of $\sim 3$~days. \citet{Youdin2011} and \citet{Beauge2013} used {\it Kepler} data to argue a similar feature was present in the radius distribution of close-in exoplanets. However, due to the preponderance of false-positives in the {\it Kepler} data at short periods \citep[e.g.][]{Morton2016}, the ``cleanliness'' of the desert in the radius-period space was not confirmed until \citet{Lundkvist2016} used a sample of exoplanets validated with asteroseismology.  \citet{Mazeh2016} further showed that the sub-Jovian desert boundaries were a function of period. The radius/mass along the lower boundary of the desert increased with increasing orbital period, while the radius/mass along the upper boundary of the desert decreased with increasing orbital period. The current {\it Kepler} exoplanet catalogues are thought to contain relatively few false positives \citep{Morton2016,Thompson2018} and the radius-period and mass-period distributions of exoplanets show the sub-Jovian desert is strikingly clean.


Atmospheric escape was identified as a possible origin of the sub-Jovain desert in the previous observational studies.  The lower boundary to the desert is well explained by the atmospheric escape from low-mass planets with large but low-mass H/He atmospheres \citep{Owen2018}. However, the origin of the upper boundary is still under debate. For example, models which use atmospheric escape calculations with large mass-loss efficiencies \citep[e.g.][]{Kurokawa2014} can explain the boundary, while those models which use atmospheric escape calculations arising from hydrodynamic models \citep{Ionov2018,Owen2018} are unable to explain the position of the boundary, instead attributing it to tidal disruption of giant planets undergoing high-eccentricity migration \citep[e.g.][]{Matsakos2016}. Therefore, it is still unclear whether the entire observed sub-Jovian desert is the evaporation desert.  

While it was possible to identify and study the evaporation desert with {\it Kepler} data, the radius precision of individual planets was around $\sim 40\%$, arising primarily from uncertainty in the stellar radius \citep{Petigura2017}. The California-Kepler-Survey used spectroscopic follow-up of exoplanet host stars to reduce the uncertainty in the stellar radius and associated planetary radii \citep{Petigura2017,Johnson2017} to $\sim 10\%$. This allowed \citet{Fulton2017} to identify a gap in the exoplanet radius and radius-period distribution shown in Figure~\ref{fig:cks}. The gap in the observed radius-period distribution is remarkably similar to the evaporation valley predicted several years earlier by \citet{Owen2013} and \citet{Lopez2013}. Further work with a sample of planets around stars for which the stellar parameters could be constrained to high precision with asterosesimology also identified the gap, but also showed that the radius of the gap declined with increasing distance from the star as is expected from the atmospheric escape model \citep{VanEylen2018}. There is still debate as to whether the observed gap in the exoplanet radius-period distribution is indeed the evaporation valley \citep[e.g.][]{Ginzburg2018,Zeng2018}. However, under the assumption that the observed gap is indeed the evaporation valley, several studies have used is properties to place constraints on the origin and evolution of the exoplanet population \citep{Owen2017,Jin2018,OMC2018}.  
	

\section{Summary}\label{sec:future}

We have come a long way in studying atmospheric escape from close-in exoplanets. This progress includes spectacular observations of both atmospheric escape occurring in nearby planetary systems and the imprints of the evolutionary consequences of atmospheric escape in the exoplanet populations. However, much of the work has still been limited in scope, primarily focusing on planets with atmospheres that are predominantly composed of Hydrogen/Helium. Even in the case of Hydroden/Helium atmospheres, the hydrodynamic simulations have not reached the point at which the launch of the outflow out to the largest scales (where the outflow starts to interact with the circumstellar environment) can be modelled. This means that even though observations of atmospheric escape occurring have been made, only crude comparisons with current models have been performed. Remedying this situation is important as many new exoplanets found by the {\it TESS} mission will also allow observational follow-up of their outflows. 

However, the field must move on from just considering escape from Hydrogen/Helium atmospheres. With the discovery of the TRAPPIST-1 system \citep{Gillon2016} it is clear that some planets may contain large amounts of water. The loss of water is critical to studies about habitability and important around lower-mass stars which have long pre-main-sequence lifetimes resulting in much higher fluxes for young planets. Much of the work on this topic performed to date has typically invoked the energy-limited model, with limited consideration as to whether it is actually applicable. Furthermore, the role of heavy elements and outgassed secondary atmospheres need to be folded into our current understanding of exoplanet evolution. For example, does a secondary atmosphere out-gassed into a Hydrogen/Helium atmosphere that is undergoing runaway atmospheric escape have a different composition to a secondary atmosphere out-gassed onto a planet that never had a large Hydrogen/Helium atmosphere to start with? As we are about to enter a period of time where obtaining spectra of close-in planets' atmospheres is possible, we need to understand atmospheric escape in order to untangle imprints of formation from escape driven evolution. 

The last decade has told us that atmospheric escape is important for driving the evolution of close-in planets; however, many open questions remain. The communities modelling efforts will need to continue to improve to keep up with the cornucopia of observations that will appear over the next decade. 


\begin{summary}[SUMMARY POINTS]
\begin{enumerate}
\item Observations of some exoplanets have detected atmospheric escape driven by hydrodynamic outflows, causing them to lose mass over time.  
\item Hydrodynamic simulations of atmospheric escape are approaching the sophistication required to compare them directly to observations.
\item Atmospheric escape sculpts sharp features into the exoplanet population that we observe today; these features have recently been detected.
\end{enumerate}
\end{summary}



\section*{DISCLOSURE STATEMENT}
The author is not aware of any affiliations, memberships, funding, or financial holdings that
might be perceived as affecting the objectivity of this review. 

\section*{ACKNOWLEDGMENTS}
JEO is supported by a Royal Society University Research Fellowship. JEO is grateful to the anonymous reviewer, David Ehrenreich, Luca Fossati and Li Zeng, for helpful comments on the manuscript. 

%

\bibliographystyle{ar-style1}
\bibliography{review_ref}

\end{document}